\documentclass[12pt]{article}
\usepackage[T1]{fontenc}
\usepackage[utf8x]{inputenc}
\usepackage{amsmath,amssymb,mathrsfs,bm}
\usepackage{subfigure}
\usepackage{graphicx}
\usepackage{color}
\begin{document}
\begin{titlepage}
\begin{center}
\begin{large}
{\bf The layered phase of anisotropic gauge theories: \\ A model for topological insulators}
\end{large}
\vskip1.0truecm
Stam Nicolis\footnote{E-mail: Stam.Nicolis@lmpt.univ-tours.fr; Stam.Nicolis@idpoisson.fr}
\vskip0.5truecm
{\sl 
Institut Denis Poisson\\
Université de Tours, Université d'Orléans, CNRS (UMR7013)\\ Parc Grandmont  37200 Tours, France
}
\end{center}
\vskip2.0truecm
\begin{abstract}
Topological insulators are materials where current does not flow through the bulk, but along the boundaries, only. They are of particular practical importance, since it is considerably more difficult, by ``conventional'' means, to affect their transport properties, than for the case of conventional materials. They are, thus, particularly robust to perturbations. 
One way to accomplish such changes is by engineering defects.  The defects that have been the most studied are domain walls; however flux compactifications can, also, work. We recall the  domain wall construction and compare it to the construction from flux compactification. 

A particular way of engineering the presence of such  defects is by introducing anisotropic couplings for the gauge fields. In this case a new phase appears, where matter is confined along layers  and local degrees of freedom cannot propagate through the bulk. It is, also, possible to take into account the ``backreaction'' of the dynamics of the gauge fields on the defects and find that a new phase, the layered phase, where, while transport of local degrees of freedom is confined to  surfaces, the topological properties can propagate through the bulk, constituting an example of anomaly flow. 

The anisotropy itself can be understood as emerging from a particular Maxwell--dilaton coupling.

\end{abstract}
\end{titlepage}
\section{Introduction}\label{intro}
While geometrical properties of matter have been understood as illustrations of general relativity, thanks to the insights it provides of how metric properties can change,  topological properties have been much harder to describe in comparable detail. First of all, classifying topological invariants is a non--trivial problem; and describing the dynamics, how topological properties can change at all, is a hard task, since, by definition, topological properties are those that do not change under smooth variations of parameters. 

A typical example where topological properties play an important role is provided by gauge theories. The terms that control these properties are, indeed, surface terms, therefore they encode ``non--local'' features of the dynamics. Therefore they aren't as easily amenable to the usual perturbative analysis and non--perturbative formulations, such as using a spacetime lattice encounter problems because the surface terms do not lead to a positive--definite Euclidian action, that can be easily sampled. 

However probing the topological properties of gauge theories, in the context of particle physics, in real experiments, already, has proven difficult; only recently have effects such as the chiral magnetic effect been shown to be accessible to experimental scrutiny in heavy--ion collisions (indeed the correct interpretation is that it is an  out--of--equilibrium effect~\cite{kharzeev2014chiral,Zubkov:2016tcp,Chernodub:2019ggz}).  On the other hand, progress in material science has led to the production of materials with properties that do highlight the relevance of surface effects, that can be associated to topological properties. 

An example of such materials is provided by the so--called ``topological insulators'' (cf.~\cite{rachel2018interacting,ludwig2015topological,hasan2010colloquium}).

Such materials are insulators, since the current does not flow through the bulk; they are topological, because current can flow along the boundaries, supported by the edge states, if such can be found. The edge states are ``protected'', if they can't mix with states in the bulk, by the topological properties of the fields--scalar and/or gauge--that serve as backgrounds. Scalar fields define ``kinks'', where the edge states are localized; gauge fields define flux backgrounds, that  serve the same purpose.

\begin{figure}[thp]
\begin{center}
\includegraphics[scale=0.5]{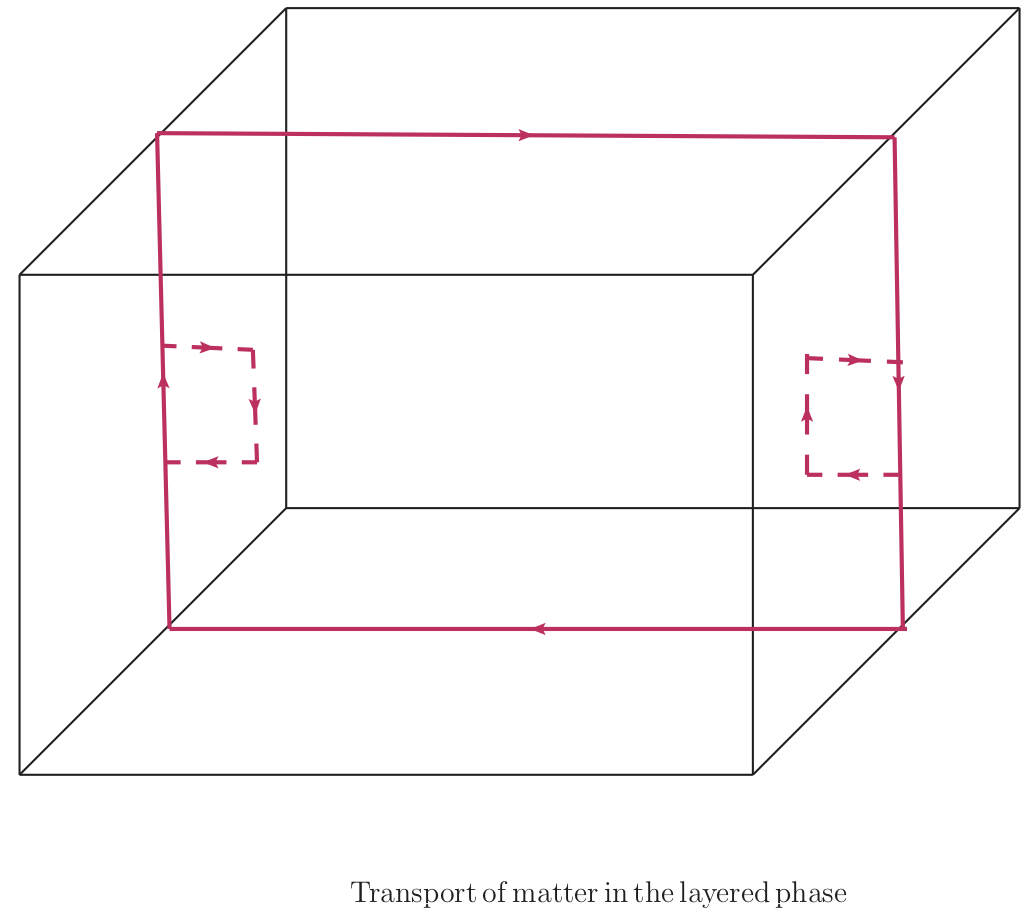}
\end{center}
\caption[]{A spacetime cartoon of a topological insulator. Current cannot flow across the bulk, only across the boundaries.}
\label{LayPhTrans}
\end{figure}

The purpose of the present note is to recall that it is precisely this scenario that is realized in anisotropic gauge theories, along the transition line between   the so--called layered phase and the bulk phases~\cite{KorthalsAltes:1993dk,Nicolis:1994xw,Hulsebos:1994pa,Nicolis:2007bz}, as indicated, schematically, in fig.~\ref{LayPhTrans}.
In the next section we shall review the construction of anisotropic lattice gauge theories and their phase diagram; we shall then provide the argument that this phase diagram provides an illustration of how it is possible to describe ways for varying the value  of the topological invariant discussed in ref.~\cite{Zubkov:2018mds,Khaidukov:2017exf}, thereby describing a ``topology--changing'' transition. 

\section{Anisotropic lattice gauge theories and the layered phase}\label{anisotlay}
The lattice regularization of gauge theories, with compact gauge group is the most direct way to compute their observables independently of any perturbative expansion. The lattice action is given by the standard, Wilson action:
\begin{equation}
\label{Wilsonaction}
\begin{array}{l}
\displaystyle
S[U]=\sum_n\,\sum_{\mu<\nu}\,\beta_{\mu\nu}\left(1-\mathrm{Re}\,\mathrm{Tr}\,U_{\mu\nu}(n)\right)\\
\displaystyle
Z = \int\,\left[\prod\,dU_\mu(n)\right]\,e^{-S[U]}
\end{array}
\end{equation}
where, however, we have taken advantage of the freedom granted by the lattice action, in assigning different couplings along different planes, $(\mu,\nu)$. Here $U_\mu(n)$ is the variable, assigned to the link $(n,n+\mu)$ and taking values in the group: $U_\mu(n)=\exp\left(\mathrm{i}\theta_a(n) T_a\right)$, where $[T_a,T_b]=\mathrm{i}f_{abc}T_c$, with $T_a$ the generators of the Lie algebra and $f_{abc}$ the structure constants. The plaquette variable, $U_{\mu\nu}(n)\equiv U_\mu(n)U_\nu(n+a\mu)U_\mu(n+a\nu)^\dagger U_\nu(n)^\dagger$ has the property of transforming under a gauge transformation, $U_\mu\to V(n)U_\mu(n)V(n)^\dagger$, 
as $U_{\mu\nu}(n)\to V(n)U_{\mu\nu}(n)V(n)^\dagger$, which implies that $\mathrm{Tr}\,U_{\mu\nu}(n)$ is a gauge--invariant quantity.  

More to the point, anisotropy of the couplings doesn't affect gauge invariance. What anisotropy does allow is to tune the couplings along different planes in independent ways, in particular to the strong coupling régime along some, while remaining in the weak coupling régime along others.  This was first remarked in ref.~\cite{Fu:1983ei,Fu:1984gj}. 

To be more specific, let us focus on the case of two couplings: $\beta$ for the planes in $d_\parallel-$dimensional subspaces and $\beta'$  for the planes that span the $d_\perp$ ``extra dimensions. The action, in this case, is given by the expression
\begin{equation}
\label{Wilsonaction1}
S[U]=\sum_n\left\{\beta\sum_{1=\mu<\nu}^{d_\parallel}\,\left(1-\mathrm{Re}\mathrm{Tr} U_{\mu\nu}\right) +
\beta'\sum_{d_\parallel+1=\mu<\nu}^{d_\parallel+d_\perp=d}\,\left(1-\mathrm{Re}\mathrm{Tr} U_{\mu\nu}\right)\right\}
\end{equation}

The question then arises, whether second order phase transitions can occur under such conditions and what are the properties of the phases, that are separated by them. The answer to the first part of the question is that such transitions can occur; the answer to the second part is that, in addition to the two phases, already known, namely the confining phase and the Coulomb phase, there exists a new phase, confining along the $d_\perp$ dimensions, and exhibiting Coulomb behavior along $d_\parallel$ dimensional ``layers''--the ``layered phase''. The order parameters are expectation values of products of the link variables, $U_\mu(n).$

To acquire some intuition about the behavior of these order parameters,  it's useful  to solve the constraints, that the link variables $U_\mu(n)$ are subject to~\cite{Nicolis:2010nh}. 

To this end, we insert in the partition function,
\begin{equation}
\label{partition_fun1}
Z[J] = \int [{\mathscr D} U] e^{-S[U]+\sum_n \mathrm{Re}(J_\mu(n) U_\mu(n))}  
\end{equation}
the expression
\begin{equation}
\label{deltafunction}
1 = \int  \left[\prod_\mathrm{links}\int d\mathrm{Re}(V_l)d\mathrm{Im}(V_l)
\delta(\mathrm{Re}(V_l)-\mathrm{Re}(U_\mu(n)))   
\delta(\mathrm{Im}(V_l)-\mathrm{Im}(U_\mu(n)))   
\right]
\end{equation}
to decouple the gauge links
\begin{equation}
\begin{array}{l}
\displaystyle
Z[J] = 
\int [{\mathscr D} U] 
\left[\prod_\mathrm{links}\int d\mathrm{Re}(V_l)d\mathrm{Im}(V_l)
\delta(\mathrm{Re}(V_l)-\mathrm{Re}(U_\mu(n)))   
\delta(\mathrm{Im}(V_l)-\mathrm{Im}(U_\mu(n)))   
\right]\\
\hskip4truecm
\times e^{-S[U]+\sum_n \mathrm{Re}(J_\mu(n) U_\mu(n))}=\\
\displaystyle
\int [{\mathscr D} U] 
\left[\prod_\mathrm{links}\int d\mathrm{Re}(V_l)d\mathrm{Im}(V_l)
\frac{d\alpha_l^R}{2\pi}\frac{d\alpha_l^I}{2\pi}
e^{\mathrm{i}\sum_l\mathrm{i}\alpha_l^R(-\mathrm{Re}(V_l)+\mathrm{Re}(U_\mu(n)))}
e^{\mathrm{i}\sum_l\mathrm{i}\alpha_l^I(-\mathrm{Im}(V_l)+\mathrm{Im}(U_\mu(n)))}
\right]\\
\hskip4truecm\times
e^{-S[U]+\sum_n \mathrm{Re}(J_\mu(n) U_\mu(n))} = \\
\displaystyle
\int
\left[\prod_\mathrm{links}\int d\mathrm{Re}(V_l)d\mathrm{Im}(V_l)
\frac{d\alpha_l^R}{2\pi}\frac{d\alpha_l^I}{2\pi}\right]\\
\hskip4truecm\times 
e^{-S[\mathrm{Re}(V_l),\mathrm{Im}(V_l)]+\sum_l(\mathrm{Re}(J_l)\mathrm{Re}(V_l)-\mathrm{Im}(J_l)\mathrm{Im}(V_l))-\mathrm{i}\sum_l\alpha_l^R\mathrm{Re}(V_l)-\mathrm{i}\sum_l\alpha_l^I\mathrm{Im}(V_l)
+\sum_l w(\alpha_l^R,\alpha_l^I)}
\end{array}
\end{equation}
where $w(\alpha_l^R,\alpha_l^I)$ contains the information about the gauge group,
\begin{equation}
\label{gauge_int}
e^{w(\alpha_l^R,\alpha_l^I)}\equiv 
\int{\mathscr D} U e^{\mathrm{i}(\alpha_l^R\mathrm{Re}(U_\mu)+\alpha_l^I\mathrm{Im}(U_\mu))}
\end{equation}
So far we have an exact transcription: we have traded the {\em constrained}
variables, $U_\mu(n)$ (that must satisfy $[\mathrm{Re}(U_\mu(n))]^2 +
[\mathrm{Im}(U_\mu(n))]^2=1$), for the {\em unconstrained} variables, $\alpha_l^R,
\alpha_l^I, \mathrm{Re}(V_l),\mathrm{Im}(V_l)$.  

The effective action seems to have acquired terms that are complex--however
the way they enter allows us to perform a ``Wick rotation''~\cite{Drouffe:1983fv}:
$\mathrm{i}\alpha_l^R\equiv
\widehat{\alpha}_l^R,\mathrm{i}\alpha_l^I\equiv\widehat{\alpha}_l^I$ and
obtain an action that is manifestly real:
\begin{equation}
\label{unconstrained}
S_\mathrm{eff}(\widehat{\alpha}_l^R,\widehat{\alpha}_l^I,\mathrm{Re}(V_l),\mathrm{Im}(V_l))
= S[\mathrm{Re}(V_l),\mathrm{Im}(V_l)] +
\sum_l\left(\widehat{\alpha}_l^R\mathrm{Re}(V_l) + \widehat{\alpha}_l^I\mathrm{Im}(V_l)\right) -
\sum_l w(\widehat{\alpha}_l^R,\widehat{\alpha}_l^I)
\end{equation}
We now look for extrema
that are uniform along the $d_\parallel$--dimensional respectively along the 
$d_\perp$ extra dimensions: $V_l\equiv v$, for links that belong in the
$d_\parallel$--dimensional subspaces and $V_l\equiv v'$ for links that ``point
out'' along the $d_\perp$ extra dimensions. Similarly $\widehat{\alpha}_l
\equiv \widehat{\alpha}$ within the $d_\parallel$ dimensional subspaces and 
$\widehat{\alpha}_l\equiv\widehat{\alpha}'$ along the $d_\perp$ extra
dimensions. 

This is a particular way of realizing the mean field approximation, that generalizes, to the case of anisotropic couplings, the 
{\em Ansatz} for isotropic couplings.

A plaquette that lies in the $d_\parallel$--dimensional subspace
makes the following contribution to the effective action
\begin{equation}
\label{dparallelplaq}
\left.\mathrm{Re}[U_{\mu\nu}(n)]\right|_{1\leq\mu<\nu\leq d_\parallel}
 = \mathrm{Re}[(v^R +
  \mathrm{i}v^I)(v^R+\mathrm{i}v^I)(v^R-\mathrm{i} v^I)(v^R-\mathrm{i} v^I)]=
([v^R]^2 + [v^I]^2)^2
\end{equation}
Similarly, a plaquette that lies in the $d_\perp$--dimensional subspace
contributes the expression
\begin{equation}
\label{dperpplaq}
\left.\mathrm{Re}[U_{\mu\nu}(n)]\right|_{d_\parallel+1\leq\mu<\nu\leq
  d_\parallel+d_\perp} = ([v'^R]^2 + [v'^I]^2)^2
\end{equation}
A plaquette that ``spans'' the subspace between two $d_\parallel$--dimensional
subspaces contributes
\begin{equation}
\label{mixedplaq}
\left.\mathrm{Re}[U_{\mu\nu}(n)]\right|_{1\leq\mu\leq d_\parallel<\nu\leq
  d_\parallel+d_\perp} = \mathrm{Re}( 
(v^R+\mathrm{i}
v^I)(v'^R+\mathrm{i}v'^I)(v^R-\mathrm{i}v^I)(v'^R-\mathrm{i}v'^I) )) = 
([v^R]^2+[v^I]^2)([v'^R]^2 + [v'^I]^2)
\end{equation}
Simple counting allows us to write down the expression for the effective
action for such uniform configurations:
\begin{equation}
\begin{array}{l}
\displaystyle
S_\mathrm{eff}[v^R,v^I,v'^R,v'^I,\widehat{\alpha}^R,\widehat{\alpha}^I,
\widehat{\alpha'}^R,\widehat{\alpha'}^I] = \\
\displaystyle
\hskip1truecm
\beta\frac{d_\parallel(d_\parallel-1)}{2}\left(1-([v^R]^2+[v^I]^2)^2\right) + 
\beta'\frac{d_\perp(d_\perp-1)}{2}\left(1-([v'^R]^2+[v'^I]^2)^2\right) + \\
\displaystyle
\hskip1truecm
\beta'd_\perp d_\parallel\left(1-([v^R]^2+[v^I]^2)([v'^R]^2 +
       [v'^I]^2)\right)+\\
\displaystyle 
\hskip1truecm
d_\parallel
\left(\widehat{\alpha}^R v^R + \widehat{\alpha}^I v^I -
w(\widehat{\alpha}^R,\widehat{\alpha}^I)\right) +
d_\perp \left(\widehat{\alpha'}^R v'^R + \widehat{\alpha'}^I v'^I -
w(\widehat{\alpha'}^R,\widehat{\alpha'}^I)\right)
\end{array}
\end{equation}
For the case of compact $U(1)$ the gauge group integral is given in terms of
elementary functions:
\begin{equation}
\label{gauge_int_U1}
e^{w(\widehat{\alpha}^R,\widehat{\alpha}^I)} = 
\int_{-\pi}^{\pi}\frac{d\theta}{2\pi} e^{\widehat{\alpha}^R\cos\theta +
  \widehat{\alpha}^I\sin\theta} = \int_{-\pi}^{\pi}\frac{d\theta}{2\pi}
e^{\sqrt{[\widehat{\alpha}^R]^2 +
    [\widehat{\alpha}^I]^2}\cos(\theta-\phi_{\widehat{\alpha}})}  \equiv 
I_0\left(\sqrt{[\widehat{\alpha}^R]^2 +
    [\widehat{\alpha}^I]^2}\right)
\end{equation}
where $I_0(\cdot)$ is the modified Bessel function. 

We notice that the group integral depends only on the length of the
``vector(s)''$(\widehat{\alpha}^R,\widehat{\alpha}^I)$--and that the plaquette
terms in the effective action depend only on the length of the ``vector(s)''
$(v^R,v^I)$. The two vectors are coupled only through their ``scalar product'',
$\widehat{\alpha}^R v^R + \widehat{\alpha}^Iv^I$, which depends on their
lengths and their
{\em relative} orientation. This means that we can {\em choose} a convenient
coordinate system in this  space \footnote{As long as the
corresponding symmetry isn't spontaneously broken. If it can be, the non--trivial minima require further study.} and we can simplify the
calculations considerably. 
We thus choose the orientations so that $v^I = 0$, $v'^I = 0$,
$\widehat{\alpha}^I=0$, $\widehat{\alpha'}^I = 0$. Indeed we easily check that 
this choice is a solution of the equations for the extrema of the effective
action. In a sense this amounts to ``choosing a gauge'' in this theory. To
simplify notation we henceforth set $v^R\equiv v$, $v'^R\equiv v'$,
$\widehat{\alpha}^R\equiv \widehat{\alpha}$, $\widehat{\alpha'}^R\equiv
\widehat{\alpha'}$. 

In this ``gauge'', therefore, the action--in the mean field approximation--takes the form
\begin{equation}
\label{gauge_fixed_action}
\begin{array}{l}
\displaystyle
S_\mathrm{eff}[v,v',\widehat{\alpha},\widehat{\alpha'}] = 
\beta\frac{d_\parallel(d_\parallel-1)}{2}\left(1-v^4\right) + 
\beta'\frac{d_\perp(d_\perp-1)}{2}\left(1-v'^4\right) + 
\beta'd_\parallel d_\perp\left(1-v^2 v'^2\right) + \\
\hskip3truecm
\displaystyle 
d_\parallel(\widehat{\alpha} v - w(\alpha)) +
d_\perp(\widehat{\alpha'}v'-w(\alpha'))
\end{array}
\end{equation}

Compactness of the gauge group implies that $w(0)=1$ and $\infty>w''(0)>0$. In
addition, $w'(0)=0$. These features may be seen to hold for compact
$U(1)$--but they hold for {\em any} compact group~\cite{Drouffe:1983fv}.

The extrema of the effective action are solutions of the equations
\begin{equation}
\label{eom}
\begin{array}{l}
\displaystyle v = dw(\widehat{\alpha})/d\widehat{\alpha}\\
\displaystyle v' = dw(\widehat{\alpha'})/d\widehat{\alpha'}\\
\displaystyle \widehat{\alpha} = 2\beta d_\parallel(d_\parallel-1)v^3+2\beta'd_\parallel
d_\perp v v'^2\\
\displaystyle \widehat{\alpha'} = 2\beta'd_\perp(d_\perp-1)v'^3+2\beta'd_\parallel v^2 v'
\end{array}
\end{equation}
in the absence of matter fields.

As $\beta$ and $\beta'$ take values from 0 to infinity, the solutions to these equations fall into three classes, that define different phases. 

The avatars of the expectation values of the Wilson loops, that are the order parameters for the different phases, in the mean field approximation, are the monomials, $v^{L_\parallel}$ (for a loop of perimeter $L_\parallel,$ that lies entirely in the $d_\parallel-$dimensional subspace), $v'^{L_\perp}$ (for a loop of perimeter $L_\perp,$ that lie entirely in the ``extra dimensions''; which is only possible if $d_\perp>1$) and $v^{L_\parallel}v'^{L_\perp}$ (for a loop of perimeter $L_\parallel+L_\perp$, that ``sticks out'' a distance $L_\perp.$ along the ``extra dimensions'').

These equations always possess the solution
$(\widehat{\alpha},\widehat{\alpha'},v,v')=(0,0,0,0)$ that corresponds to the
confining phase--since the matter fields, that are associated to sites of the lattice, can't propagate at all under such circumstances, since the gauge field is necessary for the propagation to be gauge invariant. 

However they also have
non-zero solutions, that depend on the values of the couplings $\beta$ and
$\beta'$. These do allow propagation of the matter degrees of freedom.

 The reason that these solutions can be understood as avatars of the description beyond the mean field approximation,  is 
that uniform configurations are only
invariant under global (constant) gauge transformations--and Elitzur's theorem~\cite{Elitzur:1975im} 
holds only if local transformations are possible. Thus they do   {\em not} represent a 
contradiction of Elitzur's theorem (which would signal that the mean field approximation is inconsistent) but, rather, a consequence of the fact that
the assumption behind it does not hold for the configurations under study. 

We thus find a solution with $(\widehat{\alpha},\widehat{\alpha'},v,v')\neq
(0,0,0,0)$, which corresponds to a $d_\parallel+d_\perp$--dimensional 
Coulomb phase (since Wilson loops with perimeter $L=L_1+L_2$ behave as $v^L$,
$v'^L$ or $v^{L_1}v'^{L_2}$). 

In fact we can find such a phase for any compact gauge group. It has always been assumed that this is an unphysical feature of the mean field approximation, for non--abelian groups. This statement is, however, incomplete. The correct statement is that non-abelian groups do possess a maximal abelian subgroup, corresponding to the Cartan subalgebra. The Coulomb phase for the case of non-abelian groups is nothing more or less than the phase in which the gauge fields take values in this subalgebra. This is an illustration of the ``maximal Abelian projection'' first discussed by 't Hooft~\cite{tHooft:1981bkw}; indeed the dual variables, $\widehat{\alpha}$ and $\widehat{\alpha'}$, that are confined, may be identified with the monopoles in that context~\cite{Chernodub:1997dr,sulejmanpasic2019abelian}. 

However there also exists a solution with 
$\widehat{\alpha}\neq 0,v\neq 0,\widehat{\alpha'}=0,v'=0$. In this phase (named
the ``layered phase'' in ref.~\cite{Fu:1983ei}) the Wilson loops show
perimeter behavior within a $d_\parallel$--dimensional subspace (since $v\neq
0$) and show confinement along the $d_\perp$ directions, since $v'=0$. 

In this phase, there isn't any ``bulk'' at all: the $d_\parallel+d_\perp$--dimensional space has
become a stack of $d_\parallel$--dimensional layers. The gauge coupling constant, $\beta,$ on any given layer, does, however,  
depend on the gauge coupling constant, $\beta',$ that describes the extra dimensions.  

Since the string tension
is infinite the layers are infinitely thin and the theory on them is
local. Corrections to the mean field approximation will make this string
tension finite--the layers will acquire a thickness, inversely proportional to
the (square root of the) string tension and the theory will display {\em
  non-local} features, if probed at such length scales. For this to be
consistent this string tension should be much larger than the tension of the
fundamental string. 

This is the way local degrees of freedom don't propagate through the--non-existent--bulk, but can only go from one layer to another through the boundaries.

 In all cases considered here the
boundary conditions are assumed to be periodic, but all dimensions are assumed
to become infinite in the continuum limit. 

Therefore this can be considered a new way of describing ``spontaneous compactification'' of extra spacetime dimensions, beyond what has been considered to date~\cite{Cremmer:1976zc,Randall:1999vf}, that builds upon the Kaluza--Klein paradigm. 

This analysis is summarized by the phase diagram of fig.~\ref{phadi5}:
\begin{figure}[thp]
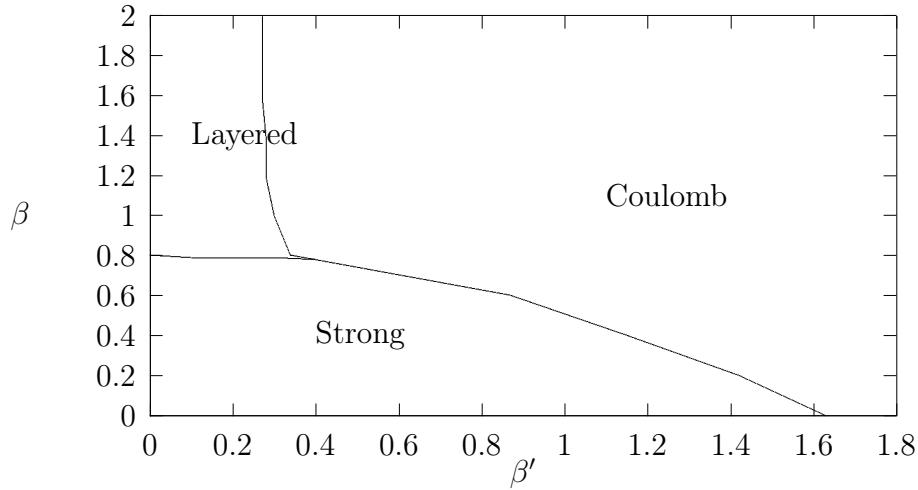

\begin{center}
\input phadi5.tex
\end{center}
\caption[]{The phase diagram of the five--dimensional $U(1)$ theory in the 
$\beta-\beta'$ plane.}
\label{phadi5}
\end{figure}
and what is of particular interest is the appearance of the layered phase, that exists for a finite interval of values for $\beta'(\beta),$ above a critical value for $\beta=\beta_\mathrm{crit}.$ 

It is interesting to try and see whether the transition from one phase to
another can become continuous. Indeed the mean field approximation to lattice
gauge theories typically predicts first order (discontinuous transitions). The
reason can be understood from the expression of the action: the plaquette
terms, in the isotropic case, are quartic in the link variables. The only
terms that can contribute to quadratic order are the ``constraint'' terms, 
$\widehat{\alpha} v - w(\widehat{\alpha})$. If we replace 
$v=dw(\widehat{\alpha})/d\alpha$ and expand to quadratic order, 
around $\alpha=0$, we find that this point corresponds to a minimum of the 
effective action,that can never become a maximum. Therefore, 
if another minimum appears for $\alpha\neq 0$, the transition is, necessarily, 
of first order. 

In the case under study here, however, there is a term in the action that {\em
  can} destabilize the confining phase in a way consistent with a continuous
transition: the term
\begin{equation}
\label{destabilization}
S_\mathrm{eff}^{\mathrm{mixed}} = \beta'd_\parallel d_\perp (1 - v^2 v'^2)
\end{equation}
is quadratic in the link variables, due to the anisotropy. And these variables
enter with a sign that allows them to destabilize the confining phase along
the $d_\perp$ directions. To see this we expand the effective action around
the solution $(\widehat{\alpha},\widehat{\alpha'}=0)$, which exists for
$\beta$ large enough and $\beta'$ small enough, within the subspace where 
$v=dw(\widehat{\alpha})/d\widehat{\alpha}$ and
$v'=dw(\widehat{\alpha'})/d\widehat{\alpha'}$. So we consider $\alpha'$ small
enough that we may expand around $\widehat{\alpha'}=0$
 to quadratic order--but we retain the exact dependence on $\alpha$. We find 
\begin{equation}
\label{destabilize}
S_\mathrm{eff}[v,v',\widehat{\alpha},\widehat{\alpha'}]\approx 
S_\mathrm{eff}[v,0,\widehat{\alpha},0] +
\widehat{\alpha}'^2 w''(0)d_\perp\left[-\beta'd_\parallel v^2(\widehat{\alpha})w''(0) + 
\frac{1}{2}\right] 
\end{equation}
This expression depends on $\beta$ implicitly, since
$\widehat{\alpha}=\widehat{\alpha}(\beta)$. If $v(\widehat{\alpha})\neq
0$--the system is in the Coulomb phase within a $d_\parallel$--dimensional
subspace--there is a line, 
\begin{equation}
\label{critical_line}
\beta'_\mathrm{crit}(\beta)
 = \frac{1}{2 d_\parallel v^2(\widehat{\alpha})w''(0)}
\end{equation} 
such that, for $\beta'<\beta'_\mathrm{crit}$ the system is in the layered
phase and for $\beta'>\beta'_\mathrm{crit}$ it is in the
$d_\parallel+d_\perp$--dimensional Coulomb phase through a continuous
transition. 
For $U(1)$, in particular, $w''(0)=1/2$ and $v(\widehat{\alpha})$
is a bounded faunction of $\widehat{\alpha}(\beta)$, that tends to 1 as
$\widehat{\alpha}(\beta)\to\infty$. In that limit, which is relevant as
$\beta\to\infty$, we obtain that $\beta'_\mathrm{crit}\to 1/d_\parallel$, a
result that is compatible with the mean field approximation, which may be
considered an expansion in $1/d_\parallel$ (and was found in another way in
ref.~\cite{Fu:1983ei}).  (Similar results can be obtained for any compact gauge group and describe configurations that take values in the Cartan subalgebra.) 
This has further interesting consequences since, many years ago,
Peskin~\cite{Peskin:1980ay} noted that at a second order phase transition point the
static quark--anti-quark potential, derived from the Wilson loop, would
display $1/R$ behavior independently of the dimensionality. 
To date an example of such a system was not
available. Anisotropic lattice gauge theories with a $U(1)$ factor could provide
such an example and it will be interesting to explore its consequences further
through Monte Carlo simulations.

\section{Coupling matter fields}\label{matter}
In the previous section we focused on the configurations of the gauge fields and the effects of the anisotropy of the couplings. In the present section we shall focus on the matter fields, in particular the fermions, studied in refs.~\cite{KorthalsAltes:1993dk,Nicolis:1994xw,Hulsebos:1994pa,Nicolis:2010jn} 
(the scalars were studied in refs.~\cite{Fu:1984gj,Dimopoulos:2000iq,Dimopoulos:2001mg} and their phase diagrams are quite intricate; the classification of the phases and the transitions that separate them hasn't been conclusively carried out. However there do exist avatars of the layered phase, in the presence of scalars.).

The coupling of matter fields to the gauge fields, on the lattice is realized (just like in the continuum) by introducing terms, linear in the link variables, {\em viz.}
\begin{equation}
\label{fermioncurrent}
S_\mathrm{int}[\psi,\overline{\psi},U]=\sum_{n,\mu}\,\left\{\overline{\psi}_n\frac{\gamma_\mu+r}{2} U_\mu(n)\psi_{n-\mu}-\overline{\psi}_{n-\mu}\frac{\gamma_\mu-r}{2}U_\mu^\dagger(n-\mu)\psi_n\right\}\equiv\sum_{n,\mu}\,J_\mu(n)^\dagger U_\mu(n)+\mathrm{h.c.}
\end{equation}
The last expression provides the definition of the (self--consistent) currents of the matter fields, that express how matter fields, that live on the sites of the lattice, couple to each other and to  the gauge fields, that live on the links in a way that makes gauge invariance an exact symmetry at finite lattice spacing (that has been taken equal to 1).
For scalar fields a similar expression can be found. 

In both cases, what is of interest is that the current, $J_\mu(n),$ since it is a quantity that couples to the gauge field, that's defined on the link, from site $n$ to site $n+\mu,$
is, also, defined on the link, as a bilinear in the fields that are defined on two nodes of the link. 

For fermions, there is an additional contribution, on the lattice,  the so--called Wilson term (proportional to the coefficient $r,$whose presence is necessary in order to ensure a correct scaling limit; it is possible to show that the value of $r$ can be taken equal to 1 and trade its tuning for that of the gauge couplings). There isn't any such term for the scalars. The reason, in a nutshell, is that the lattice introduces a spurious degeneracy for fermions, due to the fact that the kinetic term contains one spacetime derivative, whereas, for bosons, no such degeneracy is introduced, since the kinetic term for bosons contains two spacetime derivatives.  While the anisotropy does lift this degeneracy, by separating the degrees of freedom along the extra dimensions, the appearance of the layered phase shows that this separation breaks down in this case and the Wilson term is necessary for ensuring a correct scaling limit~\cite{KorthalsAltes:1993dk}.

An interesting point to keep in mind is that, upon solving the constraints, the $U_\mu(n)$ is replaced by $V,$ which, in the mean field approximation, takes the values $v$ or $v'.$ Ordinary matter fields don't couple to the $\alpha$ or $\alpha'$ variables, that would describe monopoles, since they're dual to the $v$ or $v'.$ This may be a way to describe them quantitatively, along the lines of ref.~\cite{qi2009inducing}.

We shall start by describing two ways the anisotropy of couplings can define backgrounds, that can support topological transport; we shall, then, include the coupling of the matter fields to these backgrounds by appropriate currents and review how the currents can be affected by the appearance of the layered phase. 

The equations of motion, in the mean field approximation, take into account the full effects of the backreaction; the currents are determined self--consistently.

The technical details are spelled out in refs.~\cite{KorthalsAltes:1993dk,Nicolis:1994xw} for fermions, that are the most relevant for applications. 

There are two kinds of defects, that can support topologically protected transport:
\begin{itemize}
\item {\em Domain walls}~\cite{KorthalsAltes:1993dk}: These are described by introducing a dependence of the mass of the fermions along the extra dimensions, as illustrated in fig.~\ref{domainwall}. 
\begin{figure}[thp]
\begin{center}
\includegraphics[scale=0.3]{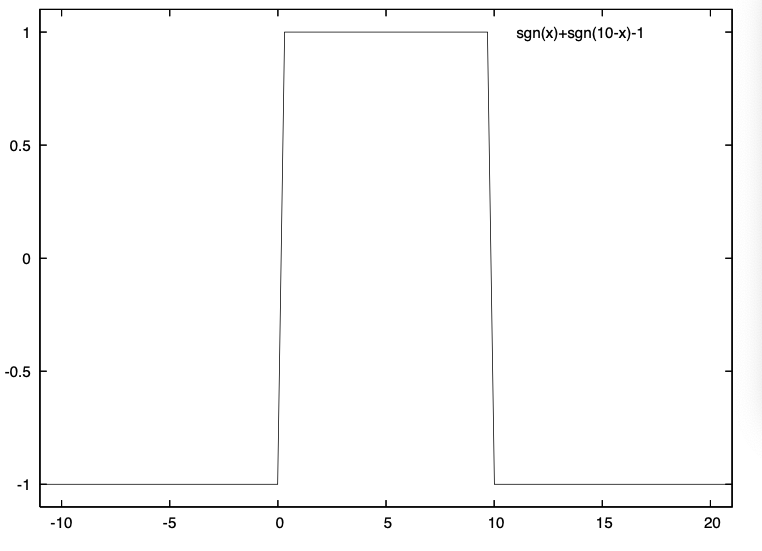}
\end{center}
\caption[]{Domain wall configuration of a kink and anti--kink, that describes the mass profile of fermions, along the extra dimension.  Fermion zero modes of a definite chirality bind to the kink and anti--kink; in the bulk the chirality isn't defined, if the bulk is odd--dimensional.}
\label{domainwall}
\end{figure}
\item {\em Flux backgrounds}: Another way of realizing  chiral fermionic zeromodes is by using fluxes. The simplest example is provided by the following gauge field configuration (cf. fig.~\ref{flux25}):
\begin{figure}[thp]
\begin{center}
\includegraphics[scale=0.3]{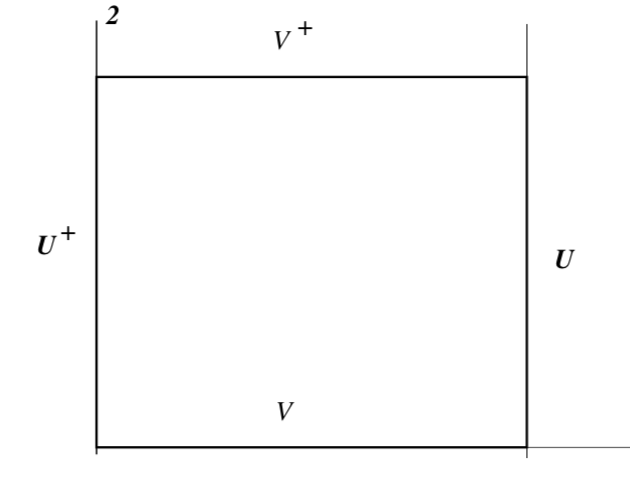}
\end{center}
\caption[]{An example of a flux background, where the links satisfy the relation $VUV^\dagger U^\dagger=e^{\mathrm{i}\Phi}$.}
\label{flux25}
\end{figure}
For constant flux, this is reminiscent of the Eguchi--Kawai models, as treated in ref.~\cite{Lebedev:1985gp}. More recently, a similar approach has been studied in refs.~\cite{Suleymanov:2018hkm,Fialkovsky:2019jeh}. 

The reason this background supports chiral fermionic zero modes is because the mean field equations of motion for the fermionic zero modes can be written as recurrences along the extra dimension in a way that's similar to that of the domain walls:
\begin{equation}
\label{fluxvacua}
\begin{array}{l}
\displaystyle
\Psi_{p,x_\perp+1}^R=-\left[M(x_\perp)+d+\sum_{\mu=1}^{d_\parallel} \cos(\Phi x_\perp+p_\mu)\right]\Psi_{p,x_\perp}^R\\
\displaystyle
\Psi_{p,x_\perp+1}^L=\left[M(x_\perp+1)+d+\sum_{\mu=1}^{d_\parallel} \cos(\Phi x_\perp+p_\mu)\right]^{-1}\Psi_{p,x_\perp}^L
\\
\end{array}
\end{equation}
where we have assumed one ``extra dimension'' along $d_\perp$. We remark that these expressions allow us to combine the domain wall profile, $M(x_\perp)$, defined by kinks/anti--kinks and the flux background(s) defined by $\Phi$. We notice that, even for a constant mass, $M(x_\perp)\equiv M_0$, these recursion relations define only one normalizable state of definite chirality, since the prefactors cannot both be less than 1. Therefore, depending on the orientation of the flux, that threads the plaquettes, only one chirality can flow along the boundary.  
\end{itemize} 
The two cases presented above show that a chiral zeromode will exist in the background of a kink/antikink pair (for the case of domain wall defects) and in the background of a flux (for the case of flux background), as long as the gauge fields aren't dynamical. 

It is  possible to compute the currents through the bulk and along the boundary, taking into account the backreaction. The mean field equations~(\ref{eom}) take the form~\cite{Hulsebos:1994pa}~\footnote{It should be stressed at this point that, though, in principle, it isn't necessary to fix the gauge for lattice gauge theories with compact gauge groups, in principle, since the integration over the volume of the gauge group just leads to the same power of this volume--the volume of the lattice--and both are finite, in practice. However, when performing numerical simulations it is necessary to fix the gauge, in order to ensure that the configurations aren't too correlated. In the above expressions we have, therefore, used axial gauge, $U_4\equiv I$.}
\begin{equation}
\label{eom1}
\begin{array}{l}
\displaystyle v = dw(\widehat{\alpha})/d\widehat{\alpha}\\
\displaystyle v' = dw(\widehat{\alpha'})/d\widehat{\alpha'}\\
\displaystyle \widehat{\alpha} = 2\beta d_\parallel(d_\parallel-1)v^3+2\beta'd_\parallel
d_\perp v v'^2 + J_\parallel\\
\displaystyle \widehat{\alpha'} = 2\beta'd_\perp(d_\perp-1)v'^3+2\beta'd_\parallel v^2 v' +J_\perp
\end{array}
\end{equation}
with $J_\parallel$ and $J_\perp$ given by the expressions
\begin{equation}
\label{currentI}
J_\parallel\equiv J_{\mu}=4\int_{-\pi}^{\pi}\frac{d^{4+D}p}{(2\pi)^{4+D}}\left[
v\sin^2 p_{\mu} + r\cos p_{\mu} W\right]\frac{1}{P}
\hspace*{1cm} \mu=1,2,3
\end{equation}
and
\begin{equation}
\label{current5I}
J_\perp\equiv J_{\nu}=4\int_{-\pi}^{\pi}\frac{d^{4+D}p}{(2\pi)^{4+D}}\left[
v'\sin^2 p_{\nu} + r\cos p_{\nu} W\right]\frac{1}{P}
\hspace*{1cm} \nu=5,\ldots,4+D
\end{equation}
where 
\begin{equation}
\label{wilson}
W\equiv M-r\left(\sum_{\lambda=1}^{3}(1-v\cos p_{\lambda})+1-\cos p_{4}+
\sum_{\lambda=5}^{4+D}(1-v'\cos p_{\lambda})\right)
\end{equation}
and
\begin{equation}
\label{denom}
P\equiv \sum_{\lambda=1}^{3}v^2\sin^2 p_{\lambda}+\sin^2 p_4 +
\sum_{\lambda=5}^{4+D}v'^2\sin^2 p_{\lambda} + W^2
\end{equation}
and, in these expressions, the values for $v$ and $v'$ are determined self--consistently, as functions of $\beta$ and $\beta'$ and, indeed, their evaluation is the most time--consuming part of the calculation, since the expressions for the currents involve multi--dimensional integrals.  

The topological invariants discussed, for instance, in~\cite{Chernodub:2015wxa} can be identified with the charge, corresponding to $J_\perp$. Its flow is therefore controlled by the properties of the latter--in particular, whether the fermion content is anomaly--free. If it is, then $J_\perp$ vanishes, since there don't exist any degrees of freedom it could couple to; if it is anomalous, then $J_\perp$ will flow in the Coulomb phase, but not in the layered phase, where the chiral zeromode is absent~\cite{KorthalsAltes:1993dk}. 

\begin{figure}[thp]
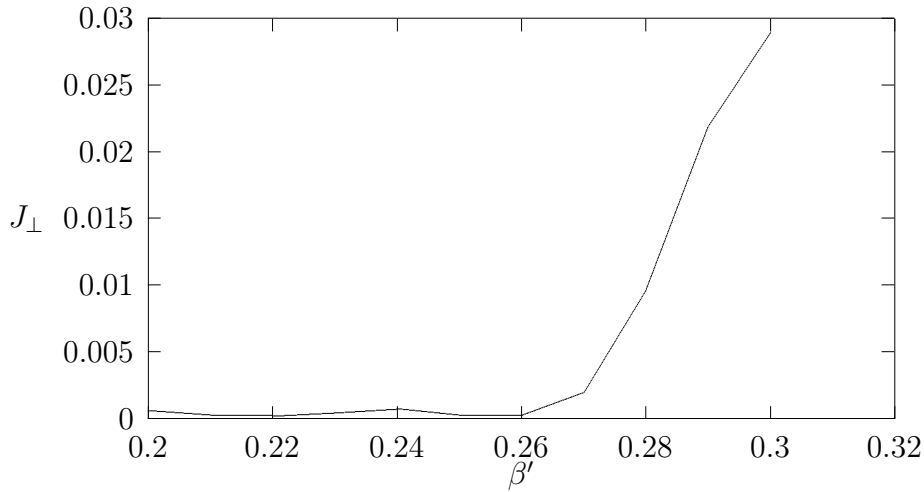

\begin{center}
\input J5.tex
\end{center}
\caption[]{The current through the bulk, as one crosses from the ``layered'' phase to the ``bulk Coulomb'' phase. In the ``layered'' phase this current vanishes, since the zeromode, that could carry it, doesn't belong to the spectrum; in the ``bulk Coulomb'' phase it does not vanish, since the theory on the boundary is anomalous.} 
\label{Jbulkanom}
\end{figure}

We can plot $J_\perp$  as a function of $\beta'$, for fixed $\beta=1.2$ in fig.~\ref{Jbulkanom}:  we cross from the ``layered'' phase to the ``bulk Coulomb'' phase. 
\section{Conclusions and outlook}\label{concl}
In this contribution we have reviewed how the salient properties of topological transport--through boundary rather than bulk degrees of freedom--can be captured by anisotropic couplings introduced in the context of lattice gauge theories, for totally different reasons. 
Introducing anisotropy in the gauge couplings leads to new ways  for controlling some quite non--trivial properties of matter coupled to gauge fields, that are squarely in the domain of strong coupling dynamics, in particular anomalous transport, through the bulk, that isn't carried by local degrees of freedom, but by excitations that are sensitive to the topology of the gauge fields. 
The transport along the boundaries is controlled by chiral excitations along the transition line from the bulk to the layered phases. These are the defining features of topological insulators, so the framework of anisotropic gauge theories seems to be the appropriate framework for describing their properties quantitatively. Within the layered phase, on the other hand, the chiral excitations, that appear, when the backreaction of the gauge fields isn't taken into account, are absent.

What hasn't received the attention it deserves is that the topology of the gauge fields can change, when these become dynamical and are coupled to dynamical matter fields--and that this can be probed within the mean field approximation~\cite{KorthalsAltes:1993dk,Nicolis:1994xw}. 
This topology change can be revealed by monitoring the current through the bulk, in odd--dimensional spacetimes, as the system passes from the layered phase to the bulk Coulomb phase. In the layered phase the current vanishes due to the absence of the chiral zeromode, whereas in the Coulomb phase it does not as the result of anomaly flow. 
(Of course if the anomaly is cancelled on the boundaries, then the current will vanish, but for different reasons. That's why the case of anomalous boundaries can provide a ``cleaner'' signal.) 

It is possible to show, by Monte Carlo simulations~\cite{Jansen:1994ym}, for the case of domain wall defects, that the mean field approximation does capture the fundamental features of the phase diagram; it is an open question to study in more detail the case of flux backgrounds in more generality and their fate, when backreaction is taken into account. 

The content of the conformal theories, defined by the continuous phase transitions between the layered phase and the bulk phases, remains to be identified, also. 
One way might be by realizing that the anisotropy of the gauge couplings can be described by a Maxwell--dilaton coupling, {\em viz.}
\begin{equation}
\label{Maxdil}
S=\int\,d^Dx\,\frac{1}{4}g^{\mu\rho}g^{\nu\sigma}F_{\mu\nu}F_{\rho\sigma}
\end{equation}
with $g^{\mu\nu}=e^\phi\delta^{\mu\nu},$ (in Euclidian signature) but where the scalar $\phi$ is ``frozen'' to its expectation value in a way that depends on the direction. This may provide hints for the continuum limit along the transition lines between layered and bulk phases and lead to interesting relations with the systems studied in ref.~\cite{Charmousis:2010zz}. In such a context, the appearance of the layered phase would seem to provide a way for stabilizing the dilaton along the extra dimensions, since its boundary is at a finite value of $\beta'.$
 
For the case of 2+1--dimensional models, that are of particular relevance for condensed matter applications (cf., for instance~\cite{cheng2019topological,devakul2020floating,devakulsubdimensional}) the transition between the bulk phases, also, may be second order. While a strong coupling expansion seems to indicate that the layered phase isn't visible in this framework~\cite{Fu:1983ei} it is, still, an open question, whether Monte Carlo simulations can detect its presence. 

Finally,  another open question pertains to the relation between the fermionic contribution and the Chern--Simons terms that define the non--trivial boundary effects of odd--dimensional spacetime geometries. It seems that using anisotropic couplings in the lattice action for the gauge fields can probe their properties. We hope to report on these issues in future work.

{\bf Acknowledgements}: It's a pleasure to thank Maxim Chernodub and Misha Zubkov for inciting me to participate in the ICNFP 2019 conference and, in particular in the Workshop on Lattice Field Theories and Condensed Matter and the other organizers of the Conference and Workshop for a wonderful event. 

I'd also like to acknowledge discussions with Bruno LeFloch on flux compactifications and to thank the referees for their comments, that contributed to improvements of the presentation. 

\bibliographystyle{utphys}
\bibliography{ICNFP2019}
\end{document}